\documentclass [18pt]{article}
\usepackage {graphicx}

\begin{document}
\textbf{On the Density Estimation by the super-parametric Method}\\\\

Yeong-Shyeong Tsai, Ying-Lin Hsu, Chung- Bow Lee, Ren-Tai Kuo and Mung-Chung
Shung

Yeong-Shyeong Tsai is a Professor of Mathematics, Theoretical Physics and
Computer Science, Department of Applied Mathematics, National Chung-Hsing
University, Taichung, Taiwan. E-mail address:ystsai@nchu.edu.tw

Ying-Lin Hsu is an Assistant Professor of Statistics, Department of Applied
Mathematics, National Chung-Hsing University, Taichung, Taiwan

Chung- Bow Lee is a Professor of Statistics, Department of Applied
Mathematics, National Chung-Hsing University, Taichung, Taiwan.

Ren-Tai Kuo is a Professor of Mathematics, Department of Applied
Mathematics, National Chung-Hsing University, Taichung, Taiwan.

Mung-Chung Shung is a Graduate Study of Computer Science,
Department of Applied Mathematics, National Chung-Hsing
University, Taichung, Taiwan.

\newpage\textbf{ABSTRACT}\\

In this paper, the approach is to study an estimator of
distribution free and to design source program which might be
useful. This distribution free estimator, super-parametric density
estimator, and its related algorithm were suggested (Tsai et al.
2007). Though we will focus on the implementation, the computer
programming, of the algorithm and strategies of choosing window
functions, the consistency of the of the estimator is studied and
the window functions such as B-spline, Bezier spline and piecewise
Bezier spline are studied as well. Since the algorithm is designed
for solving the optimization of likelihood function, there is a
set of nonlinear equations with a large number of variables. The
numerical results show that algorithm is very powerful and
effective in the sense of mathematics, that is, the iteration
procedures converge and the rate of convergence is very fast.
Though it is not main purpose to study the consistency of the
estimator, the approach in this paper to attain the consistency is
straightforward and comprehensive. From the numerical examples,
the reader can find how to use this new theory and new methods of
density estimation. The fortran source programs are appended in
this paper.\\

\textbf{1. INTRODUCTION}\\

Though it might not be a nice approach to learn statistics from
the application, the authors started to learn statistics from
computer science. There are good surveys of density estimation in
the textbook (Duda and Hart 1973). The authors studied
nonparametric density estimation, Parzen-window, from this
textbook and penalized likelihood method from papers (Good and
Gaskins 1971,1980). We think that parametric approach is well
established method. Therefore, we try to combine the theories and
techniques of both nonparametric and parametric approaches. There
are two problems must be solved. The first one is the consistency
of the estimator and the second one is the nonlinear optimization.
Though it is a hard work to encode and to debug a source program,
it is worthwhile to try a new theory and a new method. Originally,
we designed a fortran source program to test the algorithm of
nonlinear mathematical programming and to test the function of
splines. Due to the powerful theorem, Bernstein polynomial and
Stone-Weierstrass theorem, the results are so good that are beyond
our imagination,
especially, the continuous case.\\

\textbf{2. PARZEN WINDOW AND SUPER-PARAMETRIC ESTIMATOR }\\

In order to solve the second problem, we model the problem by intuition. Let
$\delta $ be Dirac delta function. Let $f$ be the density function. Clearly,
it is that

\[
f(x) = \int {\delta (x - t)f(t)dt} .
\]

Here, $f$ will be estimated by observations$x_1 $,$x_2 $,$x_3
$,\ldots ,$x_m $. The integration is replaced by summation. Let
$\widetilde{f}$ be the estimator of $f$. Let $\widetilde{f}(x) =
\sum\limits_{i = 1}^n {c_i } \varphi _i (x)$ where $\varphi _i $
are window functions,$0 \le \varphi _i (x)$, $\int {\varphi _i }
(x)dx = p_i $ and $p_i < \infty $. Let $l = \prod\limits_{j = 1}^m
{\widetilde{f}(x_j )} $ be the likelihood function. Now the
problem is to maximize $l$ subjected to the
constraints$\sum\limits_{i = 1}^n {p_i c_i = 1} $ and $0 \le c_i
$, $i = 1,2,...,n$. Mathematically, since $c_i $ are going to be
determined, if we redefine $\overline f (x) = \sum\limits_{i =
1}^n {(c_i / p_i )\varphi _i (x)} $ , then the constraints become
$\sum\limits_{i = 1}^n {c_i = 1} $ and $0 \le c_i $,$ $ $i =
1,2,...,n$. Naturally, this estimator is called super-parametric
estimator. We have noticed that most nonparametric documents give
the note: If the density function is the (linear) combinations of
window functions, then Dirac delta functions shall be obtained
when maximum likelihood estimator is applied and hence undesirable
roughness will be introduced. We think, by choosing window
function carefully, the roughness can be avoided. Before we
discuss how to choose the window function, we will quote some
results of nonlinear optimizations designed (Tsai et al. 2007).\\

\textbf{3. THE ITERATION PROCEDURES}\\

Let $\overline f (x) = \sum\limits_{i = 1}^n {u_i v_i } \varphi _i
(x)$, where $ $$\varphi _i $ are the window functions. Let
$\overline l = \prod\limits_{j = 1}^m {\overline f (x_j )} $ be
the likelihood function. Now the problem is to maximize $\overline
l $ subjected to the constraints

\begin{equation}
\label{eq1}
\sum\limits_{i = 1}^n {u_i u_i } = r.
\end{equation}

\begin{equation}
\label{eq2}
\sum\limits_{i = 1}^n {v_i v_i } = r.
\end{equation}

Let $l$ and $\overline l $ be the likelihood functions defined above. Let
$A$ be the set of all $l$. Let $\overline A $ be the set of all $\overline l
$. It is obvious that $A \subseteq \overline A $. Therefore, the maximum of
$A$ is less than or equal to that of $\overline A $. It was shown that the
extreme points of $\overline l $ should be located at the points such that
$u_i = v_i $, $i = 1,2,...,n$. Therefore, the problem to maximize $l$
subjected to its constraint is equivalent to that of maximizing $\overline l
$ subjected to the constraints (\ref{eq1}) and (\ref{eq2}). Instead of solving the problem
directly, the iteration procedures are constructed.

The procedures are:

\textbf{Step (i)}. \textit{Initialize the procedure by setting }$k
= 1$\textit{ and }$v_i^k = \sqrt {r / n} $, $ i = 1,2,...,n.$

\textbf{Step (ii)}\textit{. maximize }$\overline l $\textit{ subjected to the constraint (\ref{eq1}). Then values of }$u_i^k ,i = 1,2,...,n$\textit{, are obtained}.

\textbf{Step (iii)}. \textit{Check the condition }$\sum\limits_{i
= }^n {u_i^k v_i^k + \varepsilon \ge r} $\textit{ is satisfied or
not, where }$\varepsilon $\textit{ is a small positive number for
controlling the termination of the procedures. If the condition is
satisfied, then stop the iteration procedures and the density
estimator, }$\overline f (x) = \sum\limits_{i = 1}^n {u_i^k v_i^k
} \varphi _i (x)$\textit{, is obtained. Otherwise, increase the
value of }$k$\textit{ by one, set }$v_i^k = \theta \sqrt {u_i^{k -
1} v_i^{k - 1} } $\textit{, here }$\theta $\textit{ is a chosen
constant for satisfying constraint (\ref{eq2}). Then go to Step
(ii) and proceed the procedures.}

It is not so easy to complete step (ii), because nonlinear
optimization is very complicated usually. Let $\psi _i (x) = v_i^k
\varphi _i (x)$. The simple notation, $\overline f (x) =
\sum\limits_{i = 1}^n {u_i } \psi _i (x)$ shall be used and the
superscript of the symbols $u_i^k $ and $v_i^k $ shall be dropped
hereafter. Let $b_{ij} = \psi _i (x_j )$. Let $D_{ij} = (r /
m)({\rm {\bf b}}_i \cdot {\rm {\bf b}}_j )$. Let ${\rm {\bf u}}$
and ${\rm {\bf b}}_j $be $n$ components vectors, where ${\rm {\bf
u}} = \left[ {u_1 ,u_2 ,..,u_n } \right]^t$ and ${\rm {\bf b}}_j =
\left[ {b_{1j} ,b_{2j} ,...,b_{nj} } \right]^t$. Then ${\rm {\bf
u}} = (r / m)\sum\limits_{j = 1}^m {\alpha _j {\rm {\bf b}}_j } $,
where $\sum\limits_{j = 1}^m {D_{kj} \alpha _k \alpha _j = 1} $,
$k = 1,2,...,m$. And hence step (ii) is executed completely if the
values of all $\alpha _k $ are found. The procedures has been
designed and studied by (Tsai et al. 2007), especially, the most
complicated step, step (ii), is studied completely. Some of them
are listed in the appendix A. Since it has been studied, we will
not discuss the details of the procedures in this paper. In order
to avoid introducing the roughness, the splines are chosen as the
window
functions.\\

\textbf{4. THE CONSISTENCY OF THE ESTIMATOR}\\

In order to make the approach more comprehensive, we prefer to use
the same mathematical notation as elementary calculus. Though the
upper case letter $X_i $ and the lower case letter $x_i $ are
associated with different meanings, we try to use lower case
letter as much as possible. If we are going to study the
estimation density function which is distribution free, we may
assume that the density function, $f$, be a measurable function
defined on its domain. Let $g_{ni} (x) = 1$ for some interval
$(a_{ni} ,b_{ni} )$ and $g_{ni} (x) = 0$ otherwise. Then there is
a sequence $f_n $ such that $\mathop {\lim }\limits_{n \to \infty
} f_n (x) = f(x)$ almost everywhere, where $f_n (x) =
\sum\limits_{i = 1}^m {\theta _i^0 g_{ni} (x)} $. If we impose
some conditions on $f$, then it might be possible that $\mathop
{\lim }\limits_{n \to \infty } f_n (x) = f(x)$ uniformly. For any
$\varepsilon > 0$, there is $f_n $ such that $\left| {f(x) -
\widehat{f}(x)} \right| \le \left| {f(x) - f_n (x)} \right| +
\left| {f_n (x) - \widehat{f}(x)} \right|$. Here $\widehat{f}$ is
an estimator of $f$. Though $f$ and $f_n $ are unknown, we can
estimate $f$ if all $g_{ni} $ are known. Let $\widehat{f}(x) =
\sum\limits_{i = 1}^m {\theta _i } g_{ni} (x)$. If $\theta _i $,
$i = 1,2,...,m$, are parameters which are going to be determined,
then this is a simple parametric problem. Roughly speaking, the
nonparametric estimator can be transferred to a parametric
estimator and hence we call it super-parametric estimator. There
are many contributors (Wald 1949) who have proved the consistency
of the maximum likelihood estimators. The only problem that we
need to solve is a nonlinear optimization problem. If the density
function,$f$, is continuous on closed interval $[a,b]$, then
Bernstein polynomials play an important role in density
estimation. Here, we assume that it is well known that there are
strong connections among the Bezier spine, Bernstein polynomial
and Stone-Weierstrass theorem. Let $f$ be the density function
which is defined and continuous on $[0,1]$. Let $\widehat{f}(x) =
\sum\limits_{i = 1}^n {\theta _i \varphi _i } (x)$ be
super-parametric estimator. Let $B_n (x) = \sum\limits_{i = 0}^n
{f(x_i )C_i^n x^i(1 - x)^{n - i}} $ be Bernstein polynomial which
is associated with the density function $f$, where $C_i^n = n! /
(i!(n - i)!)$ and $x_i = i / n$. Let $\varphi _i (x) = N_i C_i^n
x^i(1 - x)^{n - i}$, here $N_i $ is a constant to make $\int
{\varphi _ i (x)dx = 1} $. Let $\theta _i^0 = f(x_i ) / N_i $.
From the property of Bernstein polynomial, we have

\begin{equation}
\label{eq3}
\left| {f(x) - \widehat{f}(x)} \right| \le \left| {f(x) - \sum\limits_{i =
0}^n {\theta _i^0 \varphi _i (x)} } \right| + \sum\limits_{i = 0}^n {\left|
{\theta _i^0 - \theta _i } \right|\left| {\varphi _i (x)} \right|}
\end{equation}

Since the first term in right hand side of the inequality can be handled, it
seems that the problem of density estimation becomes a problem of curve
fitting. Of course, it is not so simple actually.

In order to make our approach more comprehensive, we use the advantages of
mathematical notations. Let

\[
S_n = \{g;g(x) = \sum\limits_{i = 0}^n {\theta _i \varphi _i
(x),\sum\limits_{i = 0}^n {\theta _i = 1,\theta _i \ge 0,x \in D\}} } .
\]

.Here $\varphi _i $ is a normalized nonnegative function, for
example, the window function obtained from Bernstein polynomials
and $D$ is the domain of functions. Let $f$ be the density
function which is going to be estimated by a set of samples. If $f
\in S_n $, we are so lucky, then the consistency of
super-parametric estimator had been proved (Wald 1949). Generally
speaking, it is impossible to infer the uncountable information of
a density function by a finite set of samples without sufficient
assumptions and strong intuition. It should be allowed to
approximate the density function by all means. Let $X_1 $, $X_2
$,\ldots , $X_m $ be independent identically distributed from a
distribution $F$ of which the density is $f$. Let $G$ be another
distribution of which the density is $g$ and $f \ne g$. Let
$A_{\sup } = \{x;f(x)g(x) \ne 0\}$. For simplicity, we assume that
the integral is taken on $A_{\sup } $. It is obvious that

\begin{equation}
\label{eq4} \int {\log (\frac{g}{f})dF < \log (\int
{\frac{g}{f}dF)} } = 0.
\end{equation}

\noindent
since the second derivative of $ - \log $ is positive and hence $ - \log $
is convex. From the law of large number, if the number $m$ is large enough,
then we have

\begin{equation}
\label{eq5}
\frac{1}{m}\sum\limits_{i = 1}^m {\log g(X_i } ) < \frac{1}{m}\sum\limits_{i
= 1}^m {\log f(X_i } ).
\end{equation}

From (\ref{eq5}), intuitively, it is a reasonable approach to
approximate $f$ by $g_0 $ if $l_0 = \prod\limits_{j = 1}^m {g_0
(x_j )} $ and $l_0 $ is the maximum of $L$, where $L = \{l;l =
\prod\limits_{i = 1}^m {g(x_i ),g \in S_n } \}$. The existences of
$g_0 $ and $l_0 $ are doubtless since the set of $x_i $ are known
and hence $l$ is a continuous function defined on compact set
$\Omega $,\newline $\Omega = \{(\theta _1 ,\theta _2 ,...,\theta
_n );\theta _i \ge 0,\sum\limits_{i = 1}^n {\theta _i = 1\}} $.
 We will not discuss the details here. The further study is
discussed in appendix B.\\

\textbf{5. B-SPLINE AND BEZIER SPLINE}\\

Though normal distribution is a good candidate for window function
(Duda and Hart 1973), we try to use the splines as window function
in this paper. Without the Taylor's series, most useful function
might not be useful. Without the power series, there would be no
special functions which are used in classic physics and quantum
mechanics. In the practical problem, the power series shall be
replaced by polynomial. Spline is a synonym of polynomial. There
were many splines which were developed in last century (Newman and
Sproull 1979, Quarteroni Sacco and Saleri 2000). B-spline is one
of the most useful splines. In computer graphic, we will call them
blending functions instead of window functions. According to the
degree of blending polynomial functions, the blending function is
denoted by the symbol $N_{i,k} $ and defined as follows:

$N_{i,1} (x) = 1$\space if $t_i \le x < t_{i + 1} $,

$N_{i,1} (x) = 0$\space otherwise. We define them recursively,

\[
N_{i,k} (x) = \frac{(x - t_i )N_{i,k - 1} (x)}{t_{i + k - 1} - t_i
} + \frac{(t_{i + k} - x)N_{i + 1,k - 1} (x)}{t_{i + k} - t_{i +
1} }.
\]

Here, we have the convention, $\raise0.7ex\hbox{$0$}
\!\mathord{\left/ {\vphantom {0
0}}\right.\kern-\nulldelimiterspace}\!\lower0.7ex\hbox{$0$} = 0$,
and the set of knot values $t_i $,

$i = 0,1,2,...,n + k $, are defined

$t_i = x_0 $\space if $i < k$.

$t_i = x_{i - k + 1} $\space if $k \le i \le n$.

$t_i = x_{n - k + 2} $\space if $i > n$.

The estimator shall be defined the linear combinations of the
blending functions, that is $\widehat{f}(x) = \sum\limits_{i =
0}^n {c_i N_{i,k} (x)} $, where $k$ will be chosen for controlling
the order of continuity. Indeed, we can find that they are the
simple window functions used in Parzen approach when $k = 1$,
though they are not centrally located. In order to use the
B-spine, the observations should be reordered such that $x_0 \le
x_1 ,..., \le x_m $. In B-spline method, we put $ m = n$.

Bezier spines are much simpler than B-spines because there are no
extra knot points in Bezier spline. We have listed them already
and they are

\[
B_{i,n} (x) = \frac{n!}{i!(n - i)!}x^i(1 - x)^{n - i}, i =
0,1,...,n, \quad 0 \le x \le 1.
\]

Since $x$ could be defined on the specified interval, the scaling
must be done for individual problem at hand. It should be
emphasized that the algorithm (supporting document 2007) requires
that each observation data $x_j $ there must be some $\varphi _i $
such that $\varphi _i (x_j ) > 0$. If the Gauss distributions,
long tail distributions, are chosen to be the window functions,
then the requirement is satisfied automatically. Since we are in
favor of splines, the requirement makes the computer programming
more complicated, especially, in B-spline. In many applications,
B-spline has more advantages than Bezier spline. Therefore, we
started encoding the program for choosing as the window functions
the B-spline. In most documents, B-splines are defined on a
parameter or parameters, more precisely, B-spline curves and
surfaces are define by one and two parameters respectively. All
the information of the B-spline that we get is the B-spline of low
order with uniform spaced knot points. In this paper, the knot
points are the observations. Of course, the set of observation
shall be sorted and assigned to be the knot points. Unlike the
Parzen-window functions, the sizes and the shapes of the window
functions are different from one to another. And this makes some
difficulties, for example, the area of some window functions might
be zero or very small numbers and hence it is impossible to
normalize the window functions in numerical computation. These
difficulties can be handled by symbolic computation of integral by
using some computer software such as Mathematica, Maple etc. From
the figures, Figure 1 and Figure 2, we can find some effects of
non-equalized window functions, especially at two end points.
Therefore, the interval on which the windows are defined is
extended in both end points.\\

\textbf{6. THE PIECEWISE BEZIER SPLINE}\\

Bezier spline with order 10, more or less, is good enough to
estimate any continuous unimodal density function since it has 11
parameters to control the curve. If it is necessary, then the
order can be increased to 30. Classification plays an important
role in many application fields, classification by the features of
different species. Samples of pattern classification are not
obtained from a single distribution. If the density is
well-defined, then it might the mixture of different densities, so
to speak. Naturally, the piecewise Bezier spline is a nice
candidate for estimator. Moreover, the consistency we have
discussed is the case of continuous density function. If the case
we study is not continuous, then we should modify it to fit the
real case. In order to implement the piecewise Bezier spline, the
domain of distribution shall be partitioned into subdomains if it
is necessary. After the domain having been partitioned, the
estimator is the sum of Bezier splines which are defined on the
subdomains. Since the data, the samples, is the set of finite
elements, the problem is to design a simple computer program to
partition domain into subdomains. The algorithm is based on
searching for tails in the middle. In order to collaborate with
fortran source program, we use the array notation instead of
subscript index. The procedures are: (P1) Sort the samples to
obtain the ordered samples, say, $x(i)$, $i = 0,1,2,...,m$. (P2)
compute the (random) intervals $t(i)$, $t(i) = x(i + 1) - x(i)$,
(P3) Find the maximum of $t(i)$, say $t(i_0 )$. (P4) Test the
condition for partitioning the domain into two subdomains. Here,
we assume that the sample size $m = 180$. The condition is
$\raise0.7ex\hbox{$m$} \!\mathord{\left/ {\vphantom {m
6}}\right.\kern-\nulldelimiterspace}\!\lower0.7ex\hbox{$6$} < i_0
< \raise0.7ex\hbox{${5m}$} \!\mathord{\left/ {\vphantom {{5m}
6}}\right.\kern-\nulldelimiterspace}\!\lower0.7ex\hbox{$6$}$ and
$30 < i_0 < (m - 30)$. If the condition is satisfied, then the
interval $[x(i_0 ),x(i_0 + 1)]$ is removed and the domain is
partitioned into to subdomains, $[x(0),x(i_0 )]$ and $[x(i_0 +
1),x(m)]$ and hence the set of $t(i)$, not the samples, is divided
into two subsets, set $t(i_0 ) = indicator$. In order to make the
program workable, the location of must be stored in an array, say
$lo(ik)$, in the fortran program. And the value of $t(i_0 )$ is
set to be indicator, say, $t(i_0 ) = - 1$ Otherwise, that is, $
i_0\ge(5m)/6$ or $ i_0\le m/6  $, the value of $t(i_0 )$ is set to
be zero. The setting must be done to avoid being reselected. (P5)
Test the condition for terminate the algorithm. The condition is
all the length between two adjacent indicators is less than, say,
$ m/6 $. If the condition is satisfied, then terminate the
algorithm. Otherwise, go to procedure (P3). After the algorithm
being executed, the domain may be one piece interval or
partitioned into
several disconnected subintervals.\\

\textbf{7. NUMERICAL EXAMPLES}\\

In this paper, there are three numerical examples : Example 1 is
an unimodal distribution and the probability density function is
exp(-x). Example 2 is bimodal and the probability density function
is defined on $[0,4]$ , $ f(x)=2/3$ when $1\le x\le 2$ ; $f(x)
=1/3$ when $3 \le x \le 4$ ; otherwise $f(x) = 0$. Example 3 is a
trimodal distribution and the probability density function is
defined on $[0,4]$ , $f(x) = 1$ when $0 \le x \le 1/2$ ; $
f(x)=1/2$ when $ 1\le x\le 3/2$ ; $ f(x)= 1/2 $ when $ 3\le x\le
7/2 $ ; otherwise $f(x) = 0$. Before starting to design the
fortran source program for testing the formulation of this paper,
we should study the character of B-spline and Bezier spline.
Figure 1 to Figure 6 are the numerical results of this paper.
There are three methods, B-spline with order of continuity 12,
Bezier spline and piecewise Bezier spline. It is meaningless to
use large number of windows and hence the number of windows is
reduced to 11 when the sample size is 30.

Though the consistency that we have studied is the distribution
with continuous density function, the numerical examples we apply
are not confined in the continuous density function. Therefore,
some figures are not good enough. But they are
acceptable.\\

\textbf{8. DISCUSSION AND CONCLUSION}\\

Comparing to the existent results (Dong and Wetes 200, Good and
Gaskins 1971, Parzen 1962 and Rosenblatt 1956), the
super-parametric approach is a method with potential. In the
continuous case, it seems that the concavity of estimator, Bezier
spline with order 10, is almost the same as that of the density
function. So far, we think it might be a coincidence since the
concavity of a function concerns with the second derivative of the
function and the likelihood function has nothing concerning the
derivative of any function. Roughly speaking, the order of Bezier
spline should be less than 10 otherwise it will violate the
spirits the piecewise polynomials. Due the high degree of global
polynomial, the adjoin properties, oscillations will be
introduced. This is a drawback of global polynomial. If global
polynomial works well, then it is not necessary to design new
splines such as cubic spline, B-spline etc., since we have had
Lagrange polynomial. In order to apply Stone-Weierstrass theorem,
the piecewise Bezier spline shall be adopted. Though Bezier spline
can be jointed by pieces in computer aid design, the joint of
Bezier spline in density estimator should be carefully treated
because the window function in both ends have only one side tails.
This will introduce the biases of samples implicitly. Therefore,we
hope that some new flexible spline should be designed and studied.

In this paper, we focus on the splines instead of general
super-parametric approach. Though there might be some better
window functions, it seems that Bezier spline is a nice candidate
for window functions of the super-parametric estimators. The
programs were designed for testing. Therefore, they are designed
by bottom up and hence they are not readable. After having been
tested, some documents were inserted in the programs and they
become readable. Therefore, we decide to attach the source
programs in this paper. It is very easy or trivial to generalize
this approach to multivariate distributions. In multivariate
distributions, the coordinates of samples shall be transformed to
principle direction axis. This can be done by diagonalizing the
covariance matrix. After the transformation, the new coordinates
of samples are projected into the axis accordingly and hence
spline of higher dimension can be constructed by taking the
product of one dimension spline on each axis (Newman and Sproull
1979).

If all the shape of window functions are the same and each one
window covers only one sample, then we get the same result as
Parzen's approach and hence the consistency has been proved. From
the figures, we find that B-spline estimator is not so good as the
others. It must be confessed that we do not use B-spline
appropriately since we just assign the samples to knot points. If
we should choose the knot points carefully the results might be
better. Due to some reason, the roughness of estimator, we do not
try to improve choosing knot points. Though the results obtained
from B-spline is not good enough it might be useful if we follow
the approach of Bayes. Like Bayes learning, we may consider
uniform distribution as priori density, the different procedures
which we use are processes of learning and the results obtained
from the piecewise Bezier as posteriori density.\\

\textbf{APPENDIX A}\\

In order to solve the nonlinear equations $\sum\limits_{j = 1}^m
{D_{kj} \alpha _k \alpha _j = 1} $, $k = 1,2,...,m$, the iteration
procedures are constructed. First, initialize the procedure by
setting $\alpha _k = \sqrt {1 / (\overline D m)} $ where
$\overline D $ is the maximum of $D_{ij} $. Then start the
iteration procedures:

\textbf{Step (a)}. \textit{Compute }$E_i = \left| {\sum\limits_{j
= 1}^m {D_{ij} \alpha _i \alpha _j - 1} } \right|$, $i=1,2,...,m$
\textit{, and }$E = \sum\limits_{i = 1}^m {E_i } .$

\textbf{Step (b)}. \textit{Test the condition whether }$E \le
\delta $\textit{ is satisfied or not, where }$\delta $\textit{ is
a small positive number for terminating the procedures. If }$E \le
\delta $\textit{, then the desired results are obtained. Compute
}$u_k $\textit{ by the identity }$u_k = (r / m)\sum\limits_{j =
1}^m {\alpha _j b_{kj} } $, $k=1,2,...,n$\textit{. And stop the
iteration. Otherwise, }

\textbf{Step (c)}, \textit{Find the largest element of the set of
all }$E_i $\textit{. Suppose that the largest element is }$E_k
$\textit{ for some }$k$\textit{. Eliminate }$E_k $\textit{ by
updating the value of }$\alpha _k $ by $\alpha ' _k $, $\alpha '_k
= ( - s + \sqrt {s^2 + 4D_{kk} } ) / 2D_{kk} $ \textit{. And go to
Step (a).}\\

\textbf{APPENDIX B}\\

Let $S_n = \{g:g(x) = \sum\limits_{i = 0}^n {\theta _i \varphi _i
(x),\sum\limits_{i = 0}^n {\theta _i = 1,\theta _i \ge 0,x \in
D\}} } $. Let $\varphi _i $ be bounded, that is, $\left| {\varphi
_i (x)} \right| < M$. Though it is impossible to get the
information of $f$ completely, we may assume that there is a
function $g$, $g \in S_n $ such that $\left| {f(x) - g(x)} \right|
< \varepsilon $. From (\ref{eq3}) we have,

\[
\int {\left| {f(x) - \widehat{f}(x)} \right|} dF \le \int {\left| {f(x) -
g(x)} \right|} dF + \int {\left| {g(x) - \widehat{f}(x)} \right|} dF
\]

\[
\int {\left| {f(x) - \widehat{f}(x)} \right|} dF \le \int {\left| {f(x) -
g(x)} \right|} dF + \int {\left| {g(x) - \widehat{f}(x)} \right|} (dF + dG -
dG)
\]

\[
\int {\left| {f(x) - \widehat{f}(x)} \right|} dF \le \varepsilon \int {dF} +
\int {\left| {g(x) - \widehat{f}(x)} \right|} dG + \int {\left| {g(x) -
\widehat{f}(x)} \right|} \left| {f(x) - g(x)} \right|dx
\]

\[
\int {\left| {f(x) - \widehat{f}(x)} \right|} dF \le \varepsilon \int {dF} +
\int {\left| {g(x) - \widehat{f}(x)} \right|} dG + \varepsilon \int {\left|
{g(x) - \widehat{f}(x)} \right|} dx
\]

\[
\int {\left| {f(x) - \widehat{f}(x)} \right|} dF \le \varepsilon \int {dF} +
\int {\left| {g(x) - \widehat{f}(x)} \right|} dG + \varepsilon \int {(\left|
{g(x)} \right| + \left| {\widehat{f}(x)} \right|)} dx
\]

\[
\int {\left| {f(x) - \widehat{f}(x)} \right|} dF \le \varepsilon \int {dF} +
\int {\left| {g(x) - \widehat{f}(x)} \right|} dG + \varepsilon (\int
{d\widehat{F}} + \int {dG} )
\]

\[
\int {\left| {f(x) - \widehat{f}(x)} \right|} dF \le 3\varepsilon + \int
{\sum\limits_{i = 0}^n {\left| {\theta _i^0 - \theta _i } \right|} } \left|
{\varphi _i (x)} \right|dG
\]

\begin{equation}
\label{eq6}
\int {\left| {f(x) - \widehat{f}(x)} \right|} dF \le 3\varepsilon +
M\sum\limits_{i = 0}^n {\int {\left| {\theta _i^0 - \theta _i } \right|} }
dG.
\end{equation}

From (\ref{eq6}), we are able to use the samples to estimate
$\theta _i^0 $ though these samples are obtained from the
distribution $F$ instead of $G$. Here, we assume that $\varepsilon
$ is very small. There is still a problem since the existence of
$g$, $\left| {f(x) - g(x)} \right| < \varepsilon $, is not unique.
We think that it is not a real problem because the problem should
be transferred to a practical problem, the global maximum of
likelihood function. The problem has been studied partially (Tsai
et al. 2007).

\textbf{REFERENCE}

[1]. Good, I. J., and Gaskins, R. A. (1971), '' Nonparametric Roughness
Penalty for Probability Densities,'' Biometrika, Vol. 58, No. 2. pp.
255-277.

[2]. Good, I. J., and Gaskins, R. A. (1980), `` Density Estimation and Bump
Hunting by the Penalized Likelihood Method Exemplified by Scattering and
Meteorite Data,'' Journal of the America Statistical Association, Vol. 75,
No. 369, pp.42-73.

[3]. Duda, R. O., and Hart, P. E. (1973), Pattern Classification and Scene
Analysis, John Wiley {\&} Sons, New York , pp. 85-91.

[4]. Newman, W. M., and Sproull, R. F. (1979), Principle of Interactive
Computer Graphics, McGraw-Hill, New York, pp. 309-331.

[5].Silverman, B. W. (1982), `` On the Estimation Of a Probability Density
Function by the Maximum Likelihood Method,'' The Annal of Statistics, Vol.
10.No. 3. pp. 795-810.

[6]. Luenberger, D. G.. (1969), Optimization By Vector Space Methods,
pp.239-265.

[7]. Tsai, Y.-S. et al.,(2007),\textbf{ ``}Application of Quantum
Theory to Super-Parametric Density Estimation,'' .http://www.arXiv.org.

[8]. Quarteroni, A., Sacco, R., and Saleri, F. (2000), Numerical
Mathematics, Springer, Berlin pp 361-375.

[9]. Wald, A. (1949),'' Note on the Consistency of the Maximum Likelihood
Estimate,'' The Annal of Mathematical Statistics, Vol. 20, No. 4

\noindent
pp. 595-601.

[10]. Dong, M. X., and Wetes, R. J-B. (2000), `` Estimating Density
Functions: a Constrained Maximum Likelihood Approach,'' Nonparametric
statistics, Vol.

12, pp. 549-595.

[11]. Rosenblatt, M. (1956), `` Remark on some Nonparametric Estimates of a
Density Function,'' Annal Mathematical Statistics, Vol. 27, pp.832-837.

[12]. Parzen, E. (1962),'' On Estimation of a Probability Density Function
and Mode,'' Annal Mathematical Statistics, Vol. 33, pp. 1065-1076.

\newpage
\begin{figure}[htbp]
\centerline{\includegraphics[width=4.52in,height=3.86in]{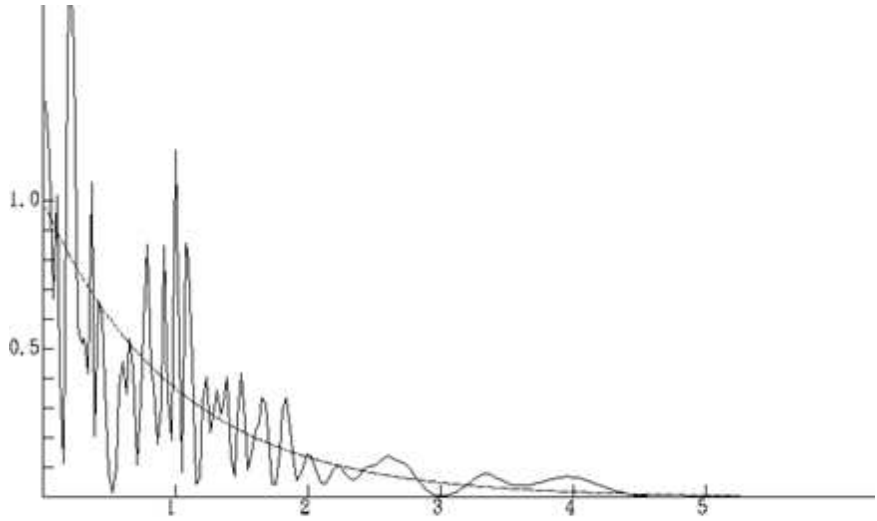}}
\caption{ In this figure, B-spline estimator is adopted to
estimate the unimodal distribution. The sample size is 180.}
\label{fig1}
\end{figure}
\begin{figure}[htbp]
\centerline{\includegraphics[width=4.42in,height=3.43in]{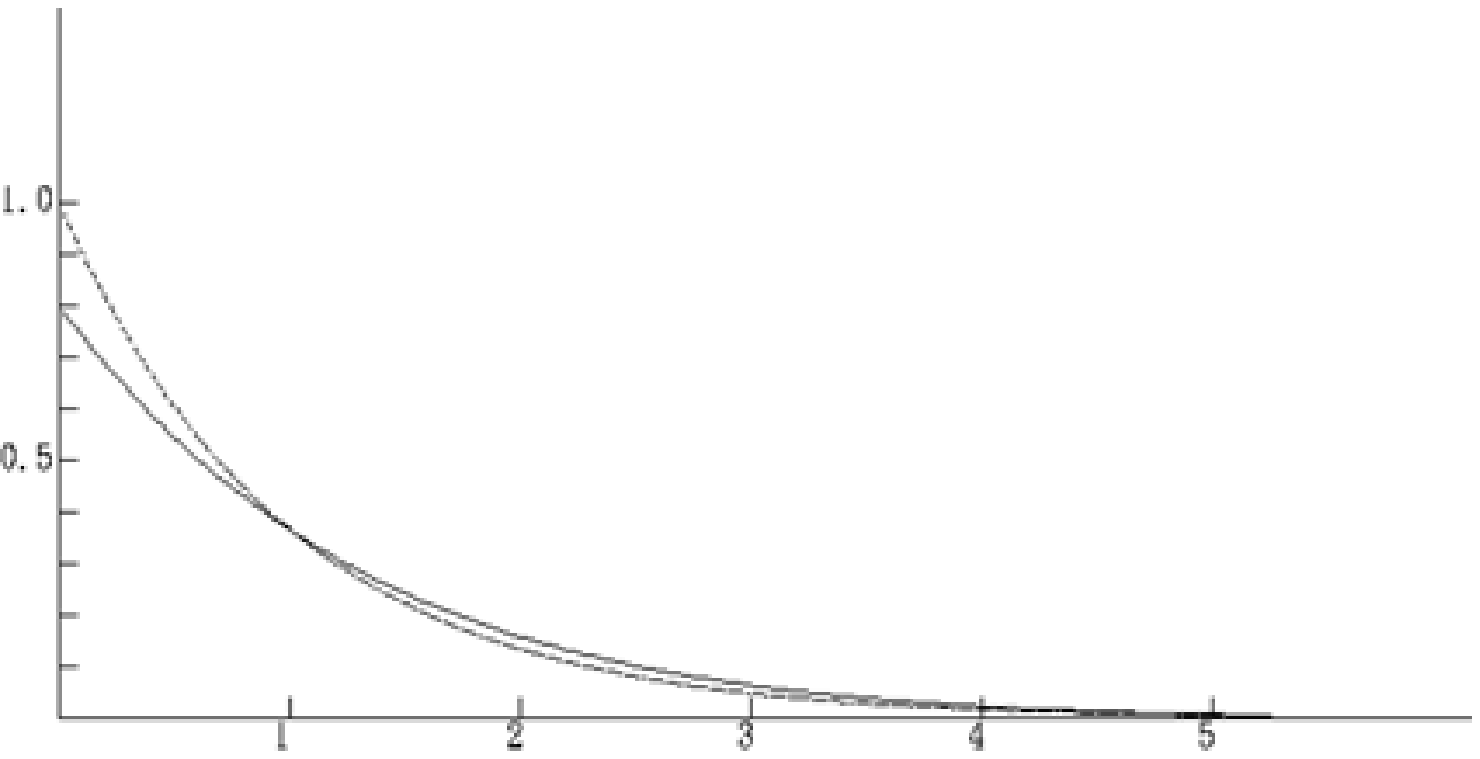}}
\caption{ In this figure, the Bezier spline estimator is adopted
to estimate the unimodal distribution. The sample size is 180.}
\label{fig2}
\end{figure}
\begin{figure}[htbp]
\centerline{\includegraphics[width=4.35in,height=3.88in]{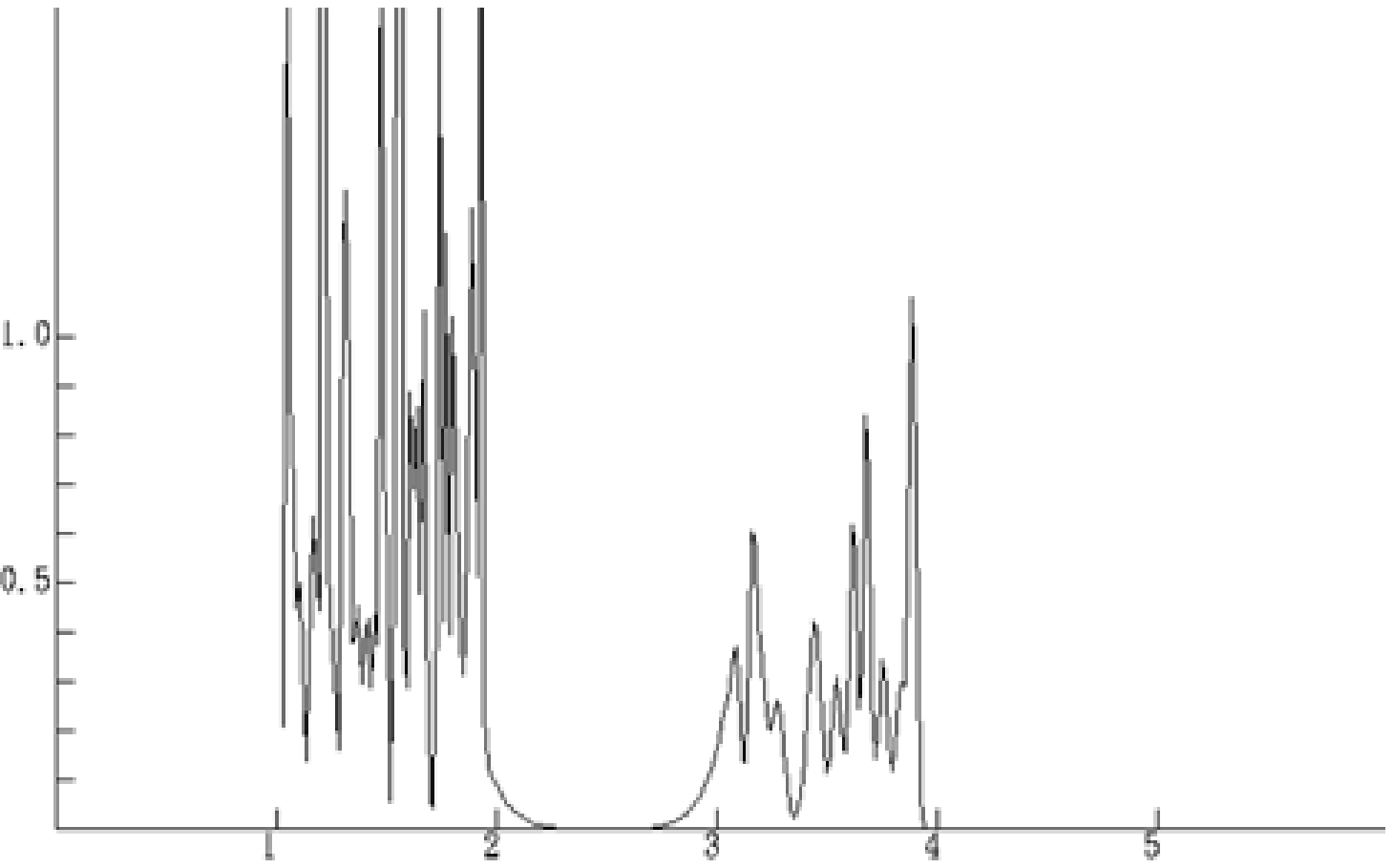}}
\caption{ In this figure, the B-spline estimator is adopted to
estimate the bimodal distribution. The sample size is 180.}
\label{fig3}
\end{figure}
\begin{figure}[htbp]
\centerline{\includegraphics[width=4.51in,height=3.55in]{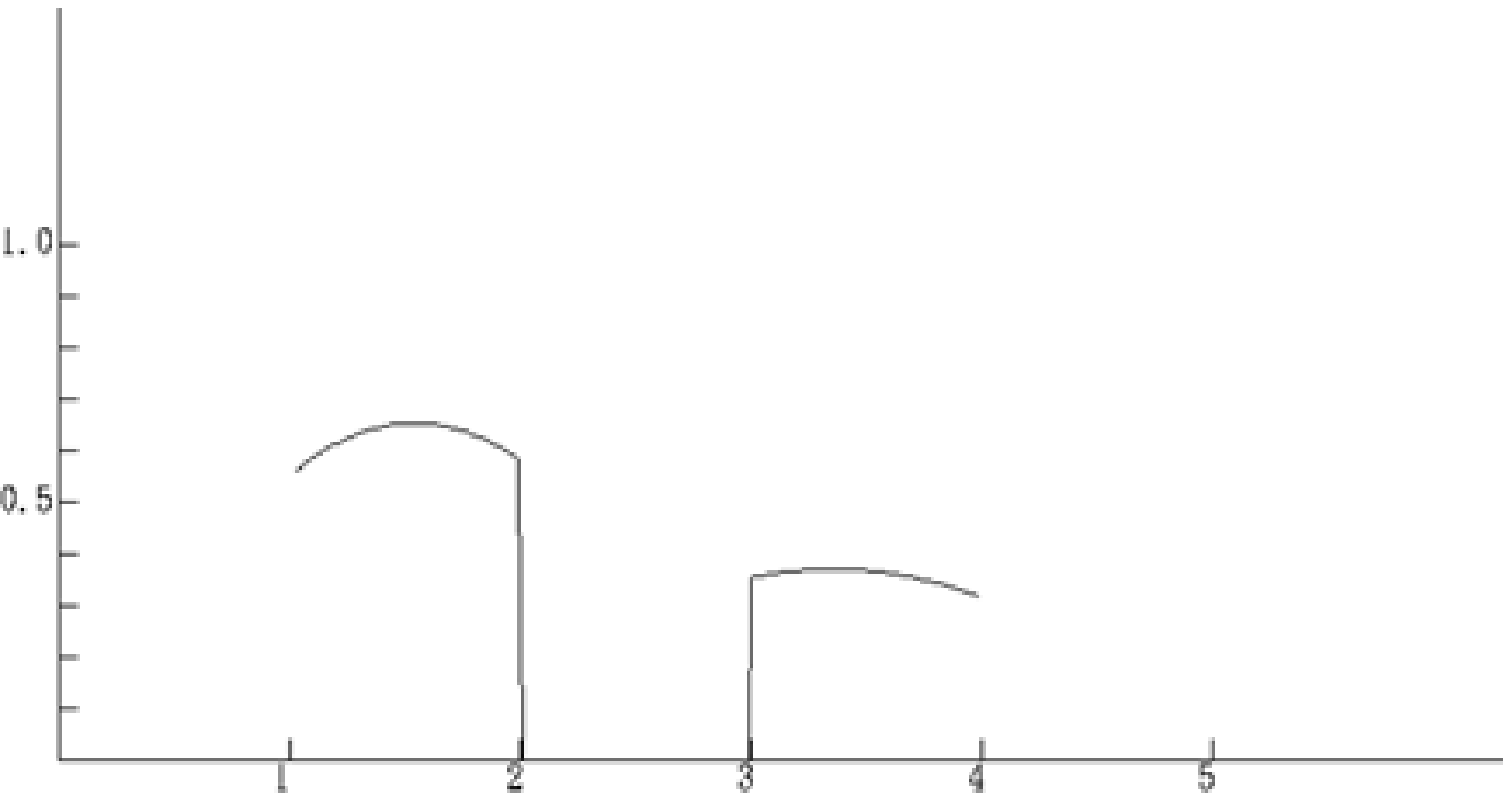}}
\caption{ In this figure, the Bezier-spline estimator is adopted
to estimate the bimodal distribution. The sample size is 180.}
\label{fig4}
\end{figure}
\begin{figure}[htbp]
\centerline{\includegraphics[width=4.44in,height=3.86in]{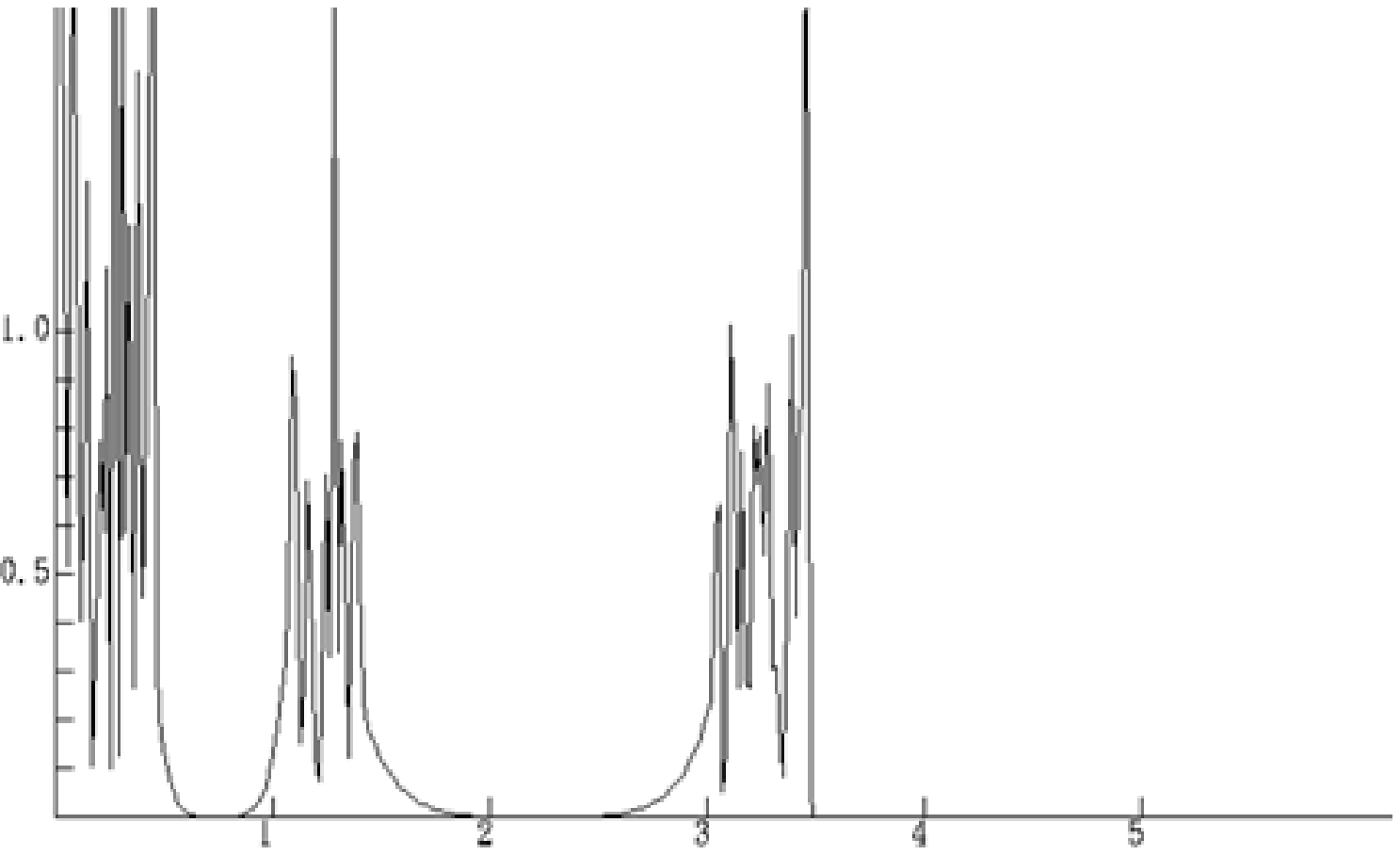}}
\caption{ In this figure, the B-spline estimator is adopted to
estimate the timodal distribution. The sample size is 180.}
\label{fig5}
\end{figure}
\begin{figure}[htbp]
\centerline{\includegraphics[width=4.36in,height=3.65in]{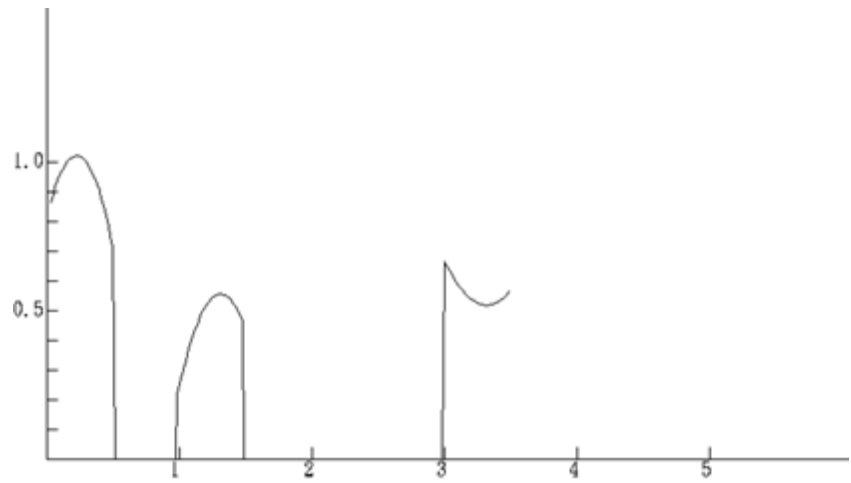}}
\caption{ In this figure, the piecewise Bezier spline estimator is
adopted to estimate the trimodal distribution. The sample size is
180.} \label{fig6}
\end{figure}

\end{document}